\documentclass{revtex4}
\usepackage{amssymb}
\usepackage{amsfonts}
\usepackage{amsmath}

\input{tcilatex}

\begin{document}

\preprint{}
\title{Revisiting the displacement operator for quantum systems with
position-dependent mass}
\author{S. Habib Mazharimousavi}
\email{habib.mazhari@emu.edu.tr}
\affiliation{Department of Physics, Eastern Mediterranean University, G. Magusa, North
Cyprus, Mersin 10 - Turkey.}

\begin{abstract}
Recently R. N. Costa Filho et al. (PRA 84, 050102(R) (2011)) have introduced
a position dependent infinitesimal translation operator which corresponds to
a position dependent linear momentum and consequently to a position
dependent effective mass quantum particle. Although there is no doubt in
novelty of the idea and the formalism, we believe that some aspects of the
quantum mechanics could be complemented in their original work. Here in this
letter first we address those points and then an alternative will be
introduced. Finally we apply the formalism for a quantum particle under a
null potential confined in a square well and the results will be compared
with those in the paper mentioned above.
\end{abstract}

\maketitle

Position dependent mass particles in non-relativistic quantum theory has
attracted attentions for the last few decades due to its application in
nuclei, impurities in crystals, $^{3}$\textbf{H}e clusters, metal clusters,
quantum liquids, semiconductor heterostructures and so on \cite{1}. More
recently, using the generalized von-Roos Hamiltonian together with the point
canonical transformation, many attempts have been made to find exact
solutions for the quantum systems with position dependent mass \cite{2}. In
a very recent attempt \cite{3}, R. N. Costa Filho et al. have approached to
the problem in a different direction. They introduced an infinitesimal
translation operator in which a well-localized state around $x$ can be
transformed to another well-localized state around $x+\left( 1+\gamma
x\right) dx$ i.e.%
\begin{equation}
\mathcal{T}_{\gamma }\left( dx\right) \left\vert x\right\rangle =\left\vert
x+dx\left( 1+\gamma x\right) \right\rangle
\end{equation}%
with all the other physical properties unchanged. Herein $\mathcal{T}%
_{\gamma }\left( dx\right) $ is the displacement operator, $\gamma $ is a
real constant with dimension (length)$^{-1}$ and $dx$ is the infinitesimal
change in the $x$ coordinate. However one should notice that, from Ref. \cite%
{3}, since 
\begin{equation}
\mathcal{T}_{\gamma }\left( dx^{\prime }\right) \mathcal{T}_{\gamma }\left(
dx^{\prime \prime }\right) =\mathcal{T}_{\gamma }\left( dx^{\prime
}+dx^{\prime \prime }+\gamma dx^{\prime }dx^{\prime \prime }\right)
\end{equation}%
and 
\begin{equation}
\exp _{q}\left( a\right) \exp _{q}\left( b\right) =\exp _{q}\left(
a+b+\left( 1-q\right) ab\right)
\end{equation}%
in which $\exp _{q}\left( a\right) $ is the $q-$exponential function, by
rewriting $\gamma =\tilde{\gamma}\left( 1-q\right) $ where $\tilde{\gamma}$
carries the unit of $\gamma $ with value one and the Tsallis entropic index $%
q$ is a real parameter, $\mathcal{T}_{\gamma }\left( dx^{\prime }\right) $
can be considered as the infinitesimal generator of the group represented by
the $q-$exponential function.

Using the standard form of the translation operator

\begin{equation}
\mathcal{T}_{\gamma }\left( dx\right) =I-\frac{i\hat{p}_{\gamma }dx}{\hslash 
}
\end{equation}%
it has been found in Ref. \cite{3} that 
\begin{equation}
\hat{p}_{\gamma }=-i\hslash \left( 1+\gamma x\right) \frac{d}{dx}
\end{equation}%
in which $\hat{p}_{\gamma }$ is the generalized generator of the translation
or the generalized linear momentum. This is easy to observe that $\hat{p}%
_{\gamma }$ is not Hermitian i.e. $\hat{p}_{\gamma }^{\dag }\neq \hat{p}%
_{\gamma }$ which implies that $\mathcal{T}_{\gamma }\left( dx\right) $ is
not unitary i.e. $\mathcal{T}_{\gamma }\left( dx\right) ^{\dag }\mathcal{T}%
_{\gamma }\left( dx\right) \neq I$. Beside the other conditions that $%
\mathcal{T}_{\gamma }\left( dx\right) $ may or may not fulfill, being
unitary looks to be more reasonable to have the normalizibility of the state
ket under the translation invariant. By following the detailed calculation
of Ref. \cite{3}, at the first look, it seems that we should sacrifice this
condition for the new form of $\mathcal{T}_{\gamma }\left( dx\right) ,$ but
by a simple manipulation this condition should not be forfeited. To see how,
here we give an alternative form for $\hat{p}_{\gamma }$ which is Hermitian.
Let's look at the details of finding the form of $\hat{p}_{\gamma }$. From
(2) one writes%
\begin{gather}
\left( I-\frac{i\hat{p}_{\gamma }\delta x}{\hslash }\right) \left\vert
\alpha \right\rangle =\mathcal{T}_{\gamma }\left( dx\right) \left\vert
\alpha \right\rangle =\int dx\mathcal{T}_{\gamma }\left( \delta x\right)
\left\vert x\right\rangle \left\langle x|\alpha \right\rangle =\int
dx\left\vert x+\delta x\left( 1+\gamma x\right) \right\rangle \left\langle
x|\alpha \right\rangle \\
=\int dx\left\vert x\right\rangle \left\langle x-\delta x\left( 1+\gamma
x\right) |\alpha \right\rangle =\int dx\left\vert x\right\rangle \left(
\left\langle x|\alpha \right\rangle -\delta x\left( 1+\gamma x\right) \frac{%
d\left\langle x|\alpha \right\rangle }{dx}\right)  \notag \\
\simeq \int dx\left\vert x\right\rangle \left( \left\langle x|\alpha
\right\rangle -\delta x\left( 1+\gamma x\right) \frac{d\left\langle x|\alpha
\right\rangle }{dx}\right) \left\vert \alpha \right\rangle +C\delta x\int
dx\left\vert x\right\rangle \left\langle x|\alpha \right\rangle  \notag \\
=\int dx\left\vert x\right\rangle \left( 1+C\delta x-\delta x\left( 1+\gamma
x\right) \frac{d}{dx}\right) \left\langle x|\alpha \right\rangle  \notag
\end{gather}%
in which $C$ is a constant to be identified later. Here we should comment
that the added term $C\left\langle x|\alpha \right\rangle \delta x$ in the
limit $\delta x\rightarrow 0$ vanishes because $\psi _{\alpha }\left(
x\right) =\left\langle x|\alpha \right\rangle $ is the wave function which
by definition is finite and square integrable. In other words this term is
negligible in comparison with the term $\left\langle x|\alpha \right\rangle $
such that $\left( 1+C\delta x\right) \left\langle x|\alpha \right\rangle
\simeq \left\langle x|\alpha \right\rangle $.

Going back to the latter equation, one finds a modified form of the
generalized linear momentum operator as follows%
\begin{equation}
\hat{p}_{\gamma }=-i\hslash \left( \left( 1+\gamma x\right) \frac{d}{dx}%
+C\right) .
\end{equation}%
To identify $C$ we impose the Hermiticity condition for $\hat{p}_{\gamma }$
which yields 
\begin{equation}
C=\frac{\gamma }{2},
\end{equation}%
and therefore 
\begin{equation}
\hat{p}_{\gamma }=-i\hslash \left( \left( 1+\gamma x\right) \frac{d}{dx}+%
\frac{\gamma }{2}\right) .
\end{equation}%
This is reasonable that $C$ vanishes when $\gamma \rightarrow 0$ because it
guarantees that $\lim_{\gamma \rightarrow 0}$ $\hat{p}_{\gamma }=-i\hslash 
\frac{d}{dx}$ which is expected. Having $\hat{p}_{\gamma }$ Hermitian
directly results a unitary translation operator $\mathcal{T}_{\gamma }\left(
dx\right) $. To draw an analogy between our formalism and Ref. \cite{3} we
rewrite 
\begin{equation}
\hat{p}_{\gamma }=-i\hslash D_{\gamma }
\end{equation}%
where 
\begin{equation}
D_{\gamma }=\left( 1+\gamma x\right) \frac{d}{dx}+\frac{\gamma }{2}
\end{equation}%
is the modified derivative in this space. Following the standard quantum
formalism for a particle with constant mass $m$ under a real potential $%
V\left( x\right) $, one finds the Schr\"{o}dinger equation 
\begin{equation}
H\psi _{\alpha }\left( x,t\right) =i\hslash \frac{\partial \psi _{\alpha
}\left( x,t\right) }{\partial t}
\end{equation}%
in which the Hamiltonian $\hat{H}$ reads%
\begin{equation}
\hat{H}=\frac{\hat{p}_{\gamma }^{2}}{2m}+V\left( x\right) .
\end{equation}%
We note here that unlike Ref. \cite{3} the new Hamiltonian is Hermitian. For
a particle of energy $E$ and a null potential the time independent Schr\"{o}%
dinger equation is given by%
\begin{equation}
-\frac{\hslash ^{2}}{2m}D_{\gamma }^{2}\phi \left( x\right) =E\phi \left(
x\right)
\end{equation}%
which after some manipulation reads%
\begin{equation}
u^{2}\phi ^{\prime \prime }\left( u\right) +au\phi ^{\prime }\left( u\right)
+b\phi \left( u\right) =0
\end{equation}%
in which $u=1+\gamma x,$ $a=3$, 
\begin{equation}
b=\frac{2m}{\hslash ^{2}\gamma ^{2}}\tilde{E}=\frac{k^{2}}{\gamma ^{2}}
\end{equation}%
and 
\begin{equation}
\tilde{E}=E+\frac{\hslash ^{2}\gamma ^{2}}{8m}.
\end{equation}%
Similar to \cite{3}, (15) is equivalent with a position dependent mass
particle with effective mass function 
\begin{equation}
m_{e}=\frac{m}{\left( 1+\gamma x\right) ^{2}}.
\end{equation}%
A general solution to (15) is given by%
\begin{equation}
\phi \left( u\right) =\frac{1}{u}\exp \left( \pm i\sqrt{\frac{k^{2}}{\gamma
^{2}}-1}\ln u\right)
\end{equation}%
in which to have a square integrable function we set $\frac{k^{2}}{\gamma
^{2}}-1>0$ or equivalently%
\begin{equation}
E>\frac{3\hslash ^{2}\gamma ^{2}}{8m}.
\end{equation}%
If we consider the particle inside an infinite well between $x=0$ and $x=L,$
the proper boundary conditions (i.e. $\phi \left( x=0\right) =0=\phi \left(
x=L\right) $) would lead to the following wave function%
\begin{equation}
\phi _{n}\left( x\right) =\left\{ 
\begin{array}{ll}
\frac{A_{n}}{\left( 1+\gamma x\right) }\sin \left( \frac{n\pi }{\ln \left(
1+\gamma L\right) }\ln \left( 1+\gamma x\right) \right) ,\text{ \ \ } & 0<x<L
\\ 
0 & \text{elsewhere}%
\end{array}%
\right.
\end{equation}%
where 
\begin{equation}
\left\vert A_{n}\right\vert ^{2}=\frac{2}{L}+2\gamma +\frac{\left( 1+\gamma
L\right) }{2n^{2}\pi ^{2}L}\ln ^{2}\left( 1+\gamma L\right) ,
\end{equation}%
\begin{equation}
k_{n}^{2}=\gamma ^{2}\left( 1+\frac{n^{2}\pi ^{2}}{\ln ^{2}\left( 1+\gamma
L\right) }\right)
\end{equation}%
and finally the energy spectrum reads as%
\begin{equation}
E_{n}=\frac{n^{2}\pi ^{2}\hslash ^{2}\gamma ^{2}}{2m\ln ^{2}\left( 1+\gamma
L\right) }+\frac{3\hslash ^{2}\gamma ^{2}}{8m}.
\end{equation}%
Easily one can show that in the limit $\gamma \rightarrow 0$ the above
results will reproduce the classical infinite potential well for a particle
with constant mass $m.$ Here we would like to compare the effect of new
configuration with those in Ref. \cite{3}. As it is clear the form of the
energy spectrum shows that the new energy is shifted up by the term $\frac{%
3\hslash ^{2}\gamma ^{2}}{8m}$ (see Fig.1 for instance). Fig. 2 displays the
density function $\left\vert \psi \right\vert ^{2}=\left\vert \phi
\right\vert ^{2}$ of the two dimensional infinite well for different values
of the quantum numbers.

Following \cite{3} we find the expectation value of the position of the
particle in one dimensional infinite well which is given by%
\begin{equation}
\left\langle x\right\rangle =\int_{0}^{L}x\left\vert \phi _{n}\left(
x\right) \right\vert ^{2}dx=\frac{\left( 1+\gamma L\right) \ln \left(
1+\gamma L\right) }{L\gamma ^{2}}\left( 1+\frac{\ln ^{2}\left( 1+\gamma
L\right) }{4\pi ^{2}n^{2}}\right) -\frac{1}{\gamma }
\end{equation}%
which implies $\lim_{\gamma \rightarrow 0}\left\langle x\right\rangle =\frac{%
L}{2}.$ Fig. 4 displays $\left\langle x\right\rangle /L$ versus $\tilde{%
\gamma}=\gamma L$ for different values of $n.$ In contrast to \cite{3}, it
is clear from Fig. 4 that $n$ plays no significant role in the general
behaviour of the diagram. Also after some manipulation one can show that the
average of the modified momentum is zero i.e., $\left\langle \hat{p}_{\gamma
}\right\rangle =0$ as it was expected.

In conclusion we add that our aim in this letter is not to criticize the
formalism given in Ref. \cite{3} and instead we try to provide a different
perspective on their new idea. In this line we have shown that how we could
introduce a linear momentum operator which is Hermitian and at the same time
matches with their formalism.

As the final point we note that although the form of the Schr\"{o}dinger
equation found in Ref. \cite{3} did not correspond with the generalized form
of the kinetic energy operator proposed by von-Roos \cite{4}, the
counterpart equation (13) in this letter is very well consistent with the
von-Roos kinetic energy operator with the ordering parameters $\alpha
=\gamma =\frac{-1}{4}$ and $\beta =\frac{-1}{2}$ \cite{5}.

\textbf{Figure captions:}

Figure 1: Relative energy spectrum of a particle in an infinite square well
and effective mass (16) versus $\tilde{\gamma}=\gamma L$ for three first
states. In this figure $E_{0}$ is the ground state energy of the particle in
the limit $\gamma \rightarrow 0$ i.e. $E_{0}=\frac{\pi ^{2}\hslash ^{2}}{%
2mL^{2}}.$ The effect of $\gamma $ is to increase ($\gamma >0$) and decrease
($\gamma <0$) the energy level.

Figure 2: The probability density of a two dimensional infinite square well
for (a) $n_{1}=1,n_{2}=1$ (b) $n_{1}=1,n_{2}=2$ (c) $n_{1}=2,n_{2}=2$ (d) $%
n_{1}=3,n_{2}=3.$ Unlike the figure reported in \cite{3} the particle is
willing to stay closer to the origin. In this figure the left side and the
right side are the top and the side view respectively and the wave functions
are normalized.

Figure 3: $\frac{\left\langle x\right\rangle }{\gamma }$ versus $\tilde{%
\gamma}=\gamma L$ of one dimensional particle in the infinite square well
for $n=1,2,3$ and $20$. Although there are slight changes between the
different cases but the general behaviors are almost the same which is in
contrast with \cite{3}.

\bigskip

\end{document}